\documentclass[12pt]{article}
\usepackage[margin=1in]{geometry} 
\usepackage{amsmath,amsthm,amssymb,todonotes}
\usepackage{multirow}
\usepackage{times}
\usepackage{listings}
\usepackage{graphicx}
\usepackage{enumerate}
\usepackage{mathrsfs}
\usepackage[colorlinks,
linkcolor=blue,
anchorcolor=blue,
citecolor=blue
]{hyperref}
\usepackage{dsfont}
\usepackage{float}
\usepackage[title]{appendix}
\usepackage{booktabs}
\usepackage{caption}
\usepackage{tikz}
\usetikzlibrary{decorations.pathreplacing}

\usepackage{chngcntr}

\usepackage{natbib}

\usepackage{setspace}
\newcommand{\R}{\mathbb{R}}

\makeatletter

\newcommand{\Rmnum}[1]{\expandafter\@slowromancap\romannumeral #1@}
\makeatother
\allowdisplaybreaks

\newtheorem{theorem}{Theorem}[section]
\newtheorem{proposition}[theorem]{Proposition}

\newtheorem{definition}[theorem]{Definition}

\numberwithin{equation}{section}

\title{Extended Relative Maximum Likelihood Updating of Choquet Beliefs}
\author{Xiaoyu Cheng\thanks{%
		Department of Managerial Economics and Decision Sciences, Kellogg School of Management, Northwestern University, Evanston, IL, USA. E-mail: xiaoyu.cheng@kellogg.northwestern.edu}}
\begin{document}
\maketitle
\begin{abstract}
\cite{cheng2019relative} proposes and characterizes Relative Maximum Likelihood (RML) updating rule when the ambiguous beliefs are represented by a set of priors. Relatedly, this note proposes and characterizes Extended RML updating rule when the ambiguous beliefs are represented by a convex capacity. Two classical updating rules for convex capacities, Dempster-Shafer \citep{Shafer1976} and Fagin-Halpern rules \citep{Fagin1990} are included as special cases of Extended RML. \\

\textit{JEL: D81, D83}

\textit{Keywords: ambiguity, updating, Choquet expected utility}  
\end{abstract}

\newpage
\section{Introduction} 
For decisions under ambiguity, the ambiguous beliefs are represented by a set of priors in the Maxmin Expected Utility (MEU) model \citep{GILBOA1989141}, and a capacity  (i.e. a non-additive probability measure) in the Choquet Expected Utility (CEU) model \citep{Schmeidler1989}. The intersection of these two models is non-empty and is given by when the set of priors in MEU is the core of the convex capacity in CEU. 

For updating ambiguous beliefs given by a set of priors, Relative Maximum Likelihood (RML) updating \citep{cheng2019relative} includes and generalizes two classical updating rules, maximum likelihood \citep{GILBOA199333} and full Bayesian \citep{pires2002rule} updating. When the set of priors is also a core of some convex capacity, maximum likelihood updating coincides with the Dempster-Shafer rule \citep{Shafer1976}. However, full Bayesian updating is in general not the same as Fagin-Halpern rule \citep{Fagin1990} noticed by \cite{ HORIE2013467}. 

More importantly, applying full Bayesian updating to the core of some convex capacity can lead to a set of posteriors that is not the core of any convex capacity. As a result, for an ex-ante preference admitting a CEU representation, the conditional preferences given by full Bayesian updating no longer admit such a representation. This is undesirable when one believes that the DM's preferences should satisfy the same set of axioms before and after receiving the information. A classical updating rule for convex capacities, Fagin-Halpern rule,  \textit{extends} the set of posteriors under full Bayesian updating exactly to its lower envelope that is also a core of some convex capacity.\footnote{Although Fagin-Halpern rule was not motivated by such an extension, it was proposed directly for capacities.}

The same problem of full Bayesian updating also applies to RML updating, as it is given by a convex combination of full Bayesian and maximum likelihood updating. Addressing this problem, this note proposes and characterizes the related \textbf{Extended RML} updating rule which is similarly given by a convex combination of Fagin-Halpern and Dempster-Shafer rules. Moreover, it is the extension of RML in the same sense as Fagin-Halpern extends full Bayesian updating. 

\section{Preliminaries}
Let $\Omega$ be the set of \textit{states} with at least three elements, endowed with a sigma-algebra $\Sigma$ of \textit{events}. Denote a generic event by $E$. Let $\Delta(\Omega)$ denote the set of all finitely additive probability measures on $(\Omega, \Sigma)$ with generic element $p$.  Let $X$ be the set of all \textit{consequences} that are simple (i.e. finite-support) lotteries over a set of prizes $Z$ and let $x$ denote a generic element of $X$. Let $\mathcal{F}$ denote the set of \textit{simple acts}, meaning that each $f \in \mathcal{F}$ is a finite-valued $\Sigma$-measurable function from $\Omega$ to $X$. With conventional abuse of notation, denote a constant act which maps all states $\omega \in \Omega$ to $x$ simply by $x$.  

The primitive is a family of preferences $\{\succsim_{E}\}_{E\in \Sigma}$ over all the acts. Let $\succsim_{\Omega} \equiv \succsim$ denote the ex-ante preference. For all the other non-empty $E\in \Sigma$, let $\succsim_{E}$ denote the conditional preference conditioning on event $E$ occurs. 

First, assume that the ex-ante preference $\succsim$ admits a Choquet Expected Utility (CEU) representation, i.e. it is represented by a non-additive measure $\nu: \Omega \rightarrow [0,1]$ and an affine utility function $u$ such that for all $f,g \in \mathcal{F}$:  
\begin{equation*}
	f \succsim  g  \Leftrightarrow  U(f) \equiv \int_{\Omega} u(f)d\nu \geq  U(g) \equiv \int_{\Omega} u(g) d\nu
\end{equation*}
where $u(f)$ denotes a random variable $Y : \Omega \rightarrow \R$ such that $Y(\omega) = u(f(\omega))$ for all $\omega \in \Omega$ and the integral is Choquet integral. Moreover, suppose the capacity is convex, i.e. for any $A,B \in \Sigma$: 
\begin{equation*}
\nu(A \cup B) + \nu(A \cap B) \geq \nu(A) + \nu(B)
\end{equation*}

It further implies the ex-ante preference admits both CEU and MEU representation. For the convex capacity $\nu$, let $C_{\nu}$ denote its core: 
\begin{equation*}
C_{\nu} = \{p \in \Delta(\Omega): p(A) \geq \nu(A), \forall A\}
\end{equation*}
Moreover, let $C^{*}_{\nu}(E)$ denote the set of distributions in $C_{\nu}$ that maximizes the probability of event $E$. 

An event $E \in \Sigma$ is $\succsim$-nonnull if $\nu(E)>0$. For every such event $E$, let $\succsim_{E}$ represent the conditional preference which also admits a CEU representation with capacity $\nu_{E}$ and the same utility function $u$: 
\begin{equation*}
	f \succsim  g  \Leftrightarrow  \int_{E} u(f)d\nu_{E} \geq  \int_{E} u(g) d\nu_{E}
\end{equation*}

The conditional preference $\succsim_{E}$ is given by Dempster-Shafer rule if it is represented by the following capacity: 
\begin{equation*}
	\nu_{E}^{DS}(A) = \frac{\nu(A\cup E^{c}) - \nu(E^{c})}{1 - \nu(E^{c})} 
\end{equation*}
Given the equivalence between Dempster-Shafer and maximum likelihood updating in this case, the core of $\nu_{E}^{DS}$ is exactly $C^{*}_{\nu}(E)$. 

The conditional preference $\succsim_{E}$ is given by Fagin-Halpern rule if it is represented by the following capacity: 
\begin{equation*}
	\nu_{E}^{FH}(A) = \frac{\nu(A\cap E)}{\nu(A\cap E) + 1 - \nu(A\cup E^{c})}
\end{equation*}

When $\nu$ is convex, both conditional capacities can also be shown to be convex. The axiomatization of these two updating rules are provided by \citet*{GILBOA199333}, \citet*{EICHBERGER2007888} and \citet{HORIE2013467}. 

Finally, define $\mathcal{F}_{E}^{2}$ to be the set of \textit{conditional binary acts}, that is for each $f \in \mathcal{F}_{E}^{2}$, there exists some $A \subseteq E$ such that $f = (b_{A}w)_{E}f$ for some $b,w\in X$ with $b \succsim w$. \cite{HORIE2013467} observes that for updating Choquet beliefs, the conditional preferences can be characterized by considering only such a subclass of acts. 

\section{Extended Relative Maximum Likelihood Updating}

\begin{definition}
	The primitive $\{\succsim_{E}\}_{E\in \Sigma}$ is represented by \textbf{Extended relative maximum likelihood updating} if there exists $\alpha \in [0,1]$ such that for all $\succsim$-nonnull $E\in \Sigma$, the conditional preference $\succsim_{E}$ admits a CEU representation with the following capacity: 
	\begin{equation*}
		\nu^{E-RML}_{E}(A) = \frac{\alpha (\nu(A\cup E^{c}) - \nu(E^{c})) + (1-\alpha)\nu(A\cap E) }{\alpha(1 - \nu(E^{c})) + (1-\alpha)(\nu(A\cap E)+1-\nu(A \cup E^{c})) }
	\end{equation*}
\end{definition}
	
	First consider the following proposition. 
	
	\begin{proposition}\label{prop1}
		When $\nu$ is a convex capacity, $\nu^{E-RML}_{E}$ is a convex capacity for any $\alpha \in [0,1]$. Moreover, for any $A \in \Sigma$, 
		\begin{equation*}
		\nu^{E-RML}_{E}(A) = \min\limits_{p \in \alpha C_{\nu}^{*}(E) + (1-\alpha) C_{\nu}} \frac{p(A\cap E)}{p(E)}. 
		\end{equation*}
	
	\end{proposition}

	This proposition establishes that $\nu^{E-RML}_{E}$ can represent a preference which admits both CEU and MEU representations. In other words, under the Extended RML updating, the conditional preferences satisfy the same set of axioms as the ex-ante axiom as desired. Moreover, it also implies that the Extended RML is the extension of RML in exactly the same sense as Fagin-Halpern extends full Bayesian updating. 
	
	Given this connection between RML and Extended RML, the axioms that characterize Extended RML are closely related to those characterize RML. Consider the following axioms: \\
	
	\textbf{Axiom CR-UO*} (Contingent Reasoning with Undershooting and Overshooting*). 
	
	For all $f\in \mathcal{F}_{E}^{2}$ and $x \in X$, if $f \sim_{E}x$ then (i) $f_{E}x \precsim x$ and (ii) $f_{E}x^{*} \succsim x_{E}x^{*}$ for all $x^{*} \succsim \max_{\omega \in E} f(\omega)$.\\
	
	\textbf{Axiom DC-CS*} (Dynamic Consistency given Conditional certainty equivalence and Sufficiently good consequences*).
	
	For all $f,g \in \mathcal{F}_{E}^{2}$ and for all $x\in X$ with $g \sim_{E}x$, if  (i) $f_{E}x \sim g_{E}x$ and (ii) $f_{E}x^{*} \sim g_{E}x^{*} $ for all $x^{*} \succsim \max_{\omega \in E} f(\omega)$, then $ f\sim_{E}g$. \\
	
	\textbf{Axiom EC*} (Event Consistency*) 
	
	For all $f,g\in \mathcal{F}_{E}^{2}$ and $x\in X$ such that $g\sim_{E_{2}} x$, if  (i) $f_{E_{1}}x \sim g_{E_{2}}x$ and (ii) $ f_{E_{1}}x_{1}^{*} \sim g_{E_{2}}x_{ 2}^{*}$ for all $x_{1}^{*} \succsim \max_{\omega \in E_{1}}f(\omega)$ and $x_{2}^{*} \succsim \max_{\omega \in E_{2}}g(\omega)$ whenever $x_{E_{1}}x_{1}^{*} \sim x_{E_{2}}x_{2}^{*}$, then $f \sim_{E_{1}} x$. \\
	
	The following theorem shows that these three axioms characterize Extended RML. 
	
	\begin{theorem}\label{thm3}
		$\{\succsim_{E}\}_{E\in \Sigma}$ is represented by Extended relative maximum likelihood updating if and only if the CR-UO*, DC-CS* and EC* axioms hold for all $\succsim$-nonnull events $E \in \Sigma$.  Furthermore, $\alpha$ is unique if there exists such an $E$ that $\nu_{E}^{DS} \neq \nu_{E}^{FH}$. 
	\end{theorem}

\section{Proof of the Results}
\subsection{Proof of Proposition \ref{prop1}}
Let $C_{\alpha}(E) =\alpha C^{*}_{\nu}(E) +  (1-\alpha )C_{\nu}$. Then any $p \in C_{\alpha}(E)$ can be written as $p = \alpha q_{1} + (1-\alpha)q_{2}$ for some $q_{1} \in C^{*}_{\nu}(E)$ and $q_{2} \in C_{\nu}$. Define a capacity $\nu'$ such that for any $A \in \Sigma$, 
\begin{equation*}
\nu'(A) \equiv \alpha \min\limits_{p\in C_{\nu}^{*}(E)}p (A) + (1-\alpha) \min\limits_{p\in C_{\nu}}p(A) 
\end{equation*}

First show that  $\nu'$ is also a convex capacity. By definition, one has the following, 

\begin{equation*}
\begin{split}
\nu'(A)  & =\alpha \min\limits_{p\in C^{*}(E)}p (A) +  (1-\alpha) \min\limits_{p\in C}p(A) \\
& = \alpha \min\limits_{p\in C^{*}(E)} [p(A\cap E) + p(A\cap E^{c})] + (1-\alpha) \nu(A)  \\
& = \alpha  \min\limits_{p\in C^{*}(E)} [p(A\cap E) + p(E^{c}) - p(A^{c}\cap E^{c})] + (1-\alpha) \nu(A) \\
& =\alpha \min\limits_{p\in C^{*}(E)} p(E^{c}) +  \alpha  \min\limits_{p\in C^{*}(E)} [p(A\cap E) - p(A^{c}\cap E^{c})] +  (1-\alpha) \nu(A)  \\
& =\alpha \min\limits_{p\in C^{*}(E)} p(E^{c}) + \alpha [\max\limits_{p\in C}p(E) - \max\limits_{p\in C} p(A^{c}\cap E) - (\max\limits_{p\in C} p(A^{c}\cup E) - \max\limits_{p\in C}p(E))] +   (1-\alpha) \nu(A) \\
& =  \alpha \nu(E^{c}) + \alpha[1- \nu(E^{c}) - (1-\nu(A\cup E^{c})) - (1- \nu(A\cap E^{c}) - (1- \nu(E^{c}))) ] + (1-\alpha) \nu(A)\\
& =  \alpha [\nu(A\cup E^{c}) + \nu(A\cap E^{c}) - \nu(E^{c})] + (1-\alpha) \nu(A) 
\end{split}
\end{equation*}
where the second equality comes from definition of $C_{\nu}$ with respect to $\nu$, the forth equality comes from the fact that $p(E^{c})$ is constant for $p\in C^{*}_{\nu}(E)$, the fifth equality is the crucial step comes from the fact that by convexity of $\nu$ there exists $p \in C^{*}_{\nu}(A^{c}\cap E) \cap C^{*}_{\nu}(E) \cap C_{\nu}^{*}(A^{c}\cup E)$ such that is exactly the one minimizes the term, the sixth equality is also by definition. 

For the fifth equality, notice that 
\begin{equation*}
\begin{split}
\min\limits_{p\in C_{\nu}^{*}(E)} [p(A\cap E) - p(A^{c}\cap E^{c})]  & = \min\limits_{p\in C_{\nu}^{*}(E)} [p(E) - p(A^{c}\cap E) - (p(A^{c}\cup E) - p(E))] \\
& = \min\limits_{p\in C_{\nu}^{*}(E)} [p(E) - p(A^{c}\cap E) - p(A^{c}\cup E) + p(E)] \\
& = 2 \max\limits_{p\in C_{\nu}} p(E) - \max\limits_{p\in C_{\nu}^{*}(E)} [p(A^{c}\cap E) + p(A^{c}\cup E)]
\end{split}
\end{equation*}
Since $A^{c}\cap E \subset E \subset A^{c}\cup E$, by convexity (comonotonicity) of $\nu$, there exists $p \in C^{*}(A^{c}\cap E) \cap C^{*}(E) \cap C^{*}(A^{c}\cup E)$, therefore the fifth equality above.\footnote{A more detailed explanation can be found in the next section.}

Then as $\nu$ is a convex capacity, it is straightforward that $\nu'$ is also convex, i.e. $\nu'(A\cup B) + \nu'(A\cap B) \geq \nu'(A) + \nu'(B)$. Then by property of Fagin-Halpern rule when the capacity is convex, one has 

\begin{equation*}
\min\limits_{p\in C_{\alpha}} \frac{p(A\cap E)}{p(E)}= \frac{\nu'(A\cap E)}{\nu'(A\cap E) + 1 - \nu'(A\cup E^{c})} 
\end{equation*}

Thus it remains to show the RHS is indeed $\nu_{E}$. By definition of $\nu'$ we have
\begin{equation*}
\begin{split}
\nu'(A\cap E) & = \alpha [\nu(A\cap E\cup E^{c}) + \nu(A\cap E\cap E^{c}) - \nu(E^{c})] + (1-\alpha) \nu(A\cap E) \\
& = \alpha[\nu(A\cup E^{c}) - \nu(E^{c})] + (1-\alpha) \nu(A\cap E) 
\end{split}
\end{equation*}
and 
\begin{equation*}
\begin{split}
\nu'(A\cup E^{c}) & =  \alpha [\nu(A\cup E^{c} \cup E^{c}) + \nu(A\cup E^{c}\cap E^{c}) - \nu(E^{c})] + (1-\alpha) \nu(A\cup E^{c}) \\
& =\alpha[\nu(A\cup E^{c}) + \nu(E^{c}) - \nu(E^{c})]  + (1-\alpha) \nu(A\cup E^{c})  = \nu(A\cup E^{c})
\end{split}
\end{equation*}

Therefore we have
\begin{equation*}
\begin{split}
\frac{\nu'(A\cap E)}{\nu'(A\cap E) + 1 - \nu'(A\cup E^{c})} & = \frac{\alpha[\nu(A\cup E^{c}) - \nu(E^{c})] + (1-\alpha) \nu(A\cap E) }{\alpha[\nu(A\cup E^{c}) - \nu(E^{c})] + (1-\alpha) \nu(A\cap E) + 1 -  \nu(A\cup E^{c})}\\
& = \frac{\alpha(\nu(A\cup E^{c}) - \nu(E^{c})) + (1-\alpha) \nu(A\cap E) }{\alpha (1- \nu(E^{c})) + (1-\alpha)( \nu(A\cap E)  + 1 - \nu(A\cup E^{c})) }
\end{split}
\end{equation*}

This proves the second statement. Also notice that, because Fagin-Halpern rule preserves convexity of a capacity, thus this also proves the first statement, i.e. $\nu^{E-RML}_{E}$ is a convex capacity.  \qed

\subsection{Proof of Theorem \ref{thm3}}
Checking necessity of the axioms is very similar as in the case of RML, thus omitted here. 

For sufficiency, the proof proceeds by the following steps: 

\underline{Step 1.} For all $\succsim$-nonnull $E\in \Sigma$, if $f\sim_{E}x$, CR-UO* implies that there exists $\alpha[E,f]$ such that for all $x^{*} \succsim \max\limits_{\omega \in E} f(\omega)$ the following holds 
\begin{equation}\label{equ5}
\alpha[E,f] U(f_{E}x^{*}) + (1-\alpha[E,f]) U(f_{E}x)  = \alpha[E,f] U(x_{E}x^{*}) + (1-\alpha[E,f]) U(x) 
\end{equation}

Since $f\in \mathcal{F}_{E}^{2}$, from now on denote it by $b_{A}w$ for some $b \succsim w$ and some $A \subseteq E$. When CR-UO* is true and $x^{*} \succsim  \max_{\omega \in E} f(\omega)$, it implies that the following inequalities hold: $f_{E}x^{*} \succsim x_{E}x^{*} \succsim x \succsim f_{E}x$. Thus, for each $x^{*}$, there exists $\alpha[E,f]$ such that equation (\ref{equ5}) holds. As the ex-ante preference is represented by the capacity $\nu$, one can further derive
\begin{equation*}
U(f_{E}x^{*}) = \int_{\Omega} u((b_{A}w)_{E}x^{*}) d\nu =u(b)(\nu(A\cup E^{c}) - \nu(E^{c})) + u(w)(1 - \nu(A\cup E^{c})) + u(x^{*})\nu(E^{c})
\end{equation*}
and 
\begin{equation*}
U(x_{E}x^{*}) = \int_{\Omega} u(x_{E}x^{*})d\nu = u(x)(1-\nu(E^{c})) +  u(x^{*})\nu(E^{c})
\end{equation*}
Notice that both expression has the common term $u(x^{*})\nu(E^{c})$, which cancels out in equation (\ref{equ5}), thus $\alpha[E,f]$ does not depend on the value of $x^{*}$. Therefore, for each $E$ and $f$, there exists $\alpha[E,f]$ such that equation (\ref{equ5}) holds for all $x^{*} \succsim \max_{\omega \in E} f(\omega)$. Furthermore, it is easy to see that $\alpha[E,f]$ is unique if either $f_{E}x \prec x$ or $f_{E}x^{*} \succ x_{E}x^{*}$ hold. \\

\underline{Step 2.} DC-CS* and EC* show that $\alpha[E,f]$ needs to be a constant across all $f\in \mathcal{F}_{E}^{2}$ and all $\succsim$-nonnull $E\in \Sigma$. Furthermore, this $\alpha$ is unique if $\nu_{E}^{DS} \neq \nu_{E}^{FH} \neq $ for some $\succsim$-nonnull $E\in \Sigma$. 

First, if $\alpha[E,f]$ is not unique for all $f$ and $E$, by result from step 1, it is true that $f_{E}x \sim x$ and $f_{E}x^{*} \sim x_{E}x^{*}$ for all $f$ and $E$. It implies that $\nu_{E}^{DS} =\nu_{E}^{FH}$ for all $\succsim$-nonnull $E$. Thus, if there exists $\nu_{E}^{FH} \neq \nu_{E}^{ML}$ for some $\succsim$-nonnull $E$, then $\alpha[E,f]$ needs to be unique for some $f$ and $E$. 

Next, to show that $\alpha[E,f]$ is the same across all $f\in \mathcal{F}_{E}^{2}$, the same argument in step 2 and 3 of the proof of Theorem 3.3 in \cite{cheng2019relative} can be directly replicated here. 

Finally, to show that $\alpha[E]$ is the same across all $\succsim$-nonnull $E\in \Sigma$, the same argument in the proof of Theorem 3.4 in \cite{cheng2019relative} can be directly replicated as well. 

Therefore, equation (\ref{equ5}) now can be written as 
\begin{equation}\label{equ6}\alpha U(f_{E}x^{*}) + (1-\alpha) U(f_{E}x)  =\alpha U(x_{E}x^{*}) +  (1-\alpha) U(x) 
\end{equation}
for some $\alpha \in [0,1]$. \\

\underline{Step 3.} Equation (\ref{equ6}) implies that the DM's conditional evaluation of any $f\in \mathcal{F}_{E}^{2}$ can be represented by 
\begin{equation*}
\int_{\Omega} u(f) d\nu_{E}
\end{equation*}
where 
\begin{equation*}
\nu_{E} = \frac{\alpha(\nu(A\cup E^{c}) - \nu(E^{c})) + (1-\alpha) \nu(A\cap E) }{\alpha(1- \nu(E^{c})) + (1-\alpha)( \nu(A\cap E)  + 1 - \nu(A\cup E^{c})) }
\end{equation*}
i.e. it is given by Extended RML with parameter $\alpha$. \\

$b_{A}w \sim_{E} x$ implies that 
\begin{equation}\label{equ1}
\int_{\Omega} u(b_{A}w)d\nu_{E}  = u(b)\nu_{E}(A) + u(w)(1-\nu_{E}(A)) = u(x)
\end{equation}
On the other hand, equation (\ref{equ6}) further implies 
\begin{equation}\label{equ2}
\begin{split}
&\alpha u(b)(\nu(A\cup E^{c}) - \nu(E^{c})) + u(w)(1 - \nu(A\cup E^{c})) + u(x^{*})\nu(E^{c})] +\\
&(1-\alpha)[u(b)\nu(A\cap E) + u(w)(1-\nu(A\cup E^{c})) + u(x)(\nu(A\cup E^{c}) - \nu(A\cap E)) ] =\\
& \alpha [u(x)(1-\nu(E^{c})) +  u(x^{*})\nu(E^{c})] + \\
&(1-\alpha) [u(x)(\nu(A\cap E)+1-\nu(A\cup E^{c})) + u(x)(\nu(A \cup E^{c}) - \nu(A\cap E)) ] \\
\end{split}
\end{equation}
since each term is evaluated according to the ex-ante preference: (the other terms are shown in step 1)
\begin{equation*}
U(f_{E}x) = \int_{\Omega} u((b_{A}w)_{E}x) d\nu = u(b)\nu(A\cap E) + u(w)(1-\nu(A\cup E^{c})) + u(x)(\nu(A\cup E^{c}) - \nu(A\cap E))   \\
\end{equation*}
where the second equality comes from the fact that $b \succsim x \succsim w$. 

Then arranging terms in equation (\ref{equ2}) by eliminating the common terms $\alpha u(x^{*})\nu(E^{c}) + (1-\alpha)u(x)(\nu(A\cup E^{c}) - \nu(A\cap E)) $ on both sides we get, 
\begin{equation}\label{equ3}
\begin{split}
&\alpha[u(b)(\nu(A\cup E^{c}) - \nu(E^{c})) + u(w)(1 - \nu(A\cup E^{c}))] +\\
&(1-\alpha)[u(b)\nu(A\cap E) + u(w)(1-\nu(A\cup E^{c}))] = \\
&\alpha u(x)(1-\nu(E^{c})) + (1-\alpha) u(x)(\nu(A\cap E)+1-\nu(A\cup E^{c})) 
\end{split}
\end{equation}
Notice that, by CR-UO* axiom, equation (\ref{equ6}) implies equation (\ref{equ3}) for all $b\succsim w$, therefore $\nu_{E}(A)$ is given by exactly the definition in the Extended RML updating with $\alpha$. \qed

\section{Core of a Convex Capacity}
In this section, I provide an equivalent characterization of a set of priors being the core of a convex capacity. 

Fix a set of distributions $C$, a sequence of events $A_{1} \subseteq A_{2} \subseteq \cdots \subseteq A_{n} \subseteq \Omega$ are called \textit{comonotonic events} if the following holds: 
\begin{equation*}
C^{*}(A_{1}) \cap C^{*}(A_{2}) \cap \cdots \cap C^{*}(A_{n}) \neq \emptyset
\end{equation*}
i.e. for the sequence of events, there exists a probability measure in $C$ such that attains the maximum likelihood of all events in the sequence \textit{simultaneously}. In other words, comonotonic events are ``separable'' according to the set of priors in the sense that whether or not one event achieves its maximum likelihood does not affect whether the events that are superset or subset of it \textit{can} achieve their maximum likelihood. 

A set $C$ is \textit{comonotonic} if all sequence of events are comonotonic events, which gives the characterization of the core of convex capacity: 

\begin{proposition}\label{prop6}
	A convex and compact set $C$ is the core of a convex capacity if and only if it is comonotonic. 
\end{proposition}

The proof of this proposition follows from Theorem 2 in \cite{shapley1971cores}.

\newpage
\bibliographystyle{econ}
\bibliography{/Users/xiaoyucheng/Dropbox/Research/References/mylibrary.bib}

\end{document}